\documentclass{aa}     
\usepackage{graphicx}

\newcommand{\Mpc}{$h^{-1}$\thinspace Mpc}

\begin{document}   

                       
\title{Las Campanas Loose Groups in the Supercluster-Void Network}

\author {M. Einasto\inst{1}, J. Jaaniste\inst{1,2}, 
J. Einasto\inst{1}, P. Hein\"am\"aki\inst{1,3}, 
V. M\"uller\inst{4}, D.L. Tucker\inst{5}}
\authorrunning{M. Einasto et al.}

\offprints{M. Einasto }

\institute{ Tartu Observatory, EE-61602 T\~oravere, Estonia
\and
Institute of Physics, Estonian Agricultural University, 
Kreutzwaldi 64, EE-51014 Tartu, Estonia
\and
Tuorla Observatory, V\"ais\"al\"antie 20, Piikki\"o, Finland 
\and
Astrophysical Institute Potsdam, An der Sternwarte 16,
D-14482 Potsdam, Germany
\and
 Fermi National Accelerator Laboratory, MS 127, PO Box 500, Batavia, 
IL 60510, USA
}

\date{ Received   2003 / Accepted ...  } 

\titlerunning{LGS}


\abstract{
We study the spatial distribution of  loose groups from the Las Campanas
Redshift  Survey, comparing it with  the  supercluster-void network
delineated by  rich clusters of galaxies.  We   use density fields and
the friends-of-friends (FoF) algorithm  to identify the members of 
superclusters of  Abell clusters among the Las Campanas loose groups. 
We find that systems of loose groups tend 
to be oriented perpendicularly to  the
line-of-sight, and discuss possible reasons for that.
We show that loose groups in  richer systems (superclusters
of Abell clusters) are themselves also  richer and more massive than
groups in systems without Abell clusters. Our results indicate
that superclusters, as high density environments, have a major role in
the formation and evolution of galaxy systems.

\keywords{cosmology: observations -- cosmology: large-scale structure
of the Universe}
}

\maketitle

\section{Introduction}
The first evidence that the large scale structure of the Universe
forms a web-like network of galaxy systems was obtained in  the 1970's  
(J\~oeveer \& Einasto \cite{joe78},  Tarenghi et al.   \cite{tar78}).
In  the supercluster-void network,  superclusters  of galaxies
with
characteristic dimensions of  up  to 100~\Mpc\ \footnote{$h$  is
the Hubble constant   in units of 100~km~s$^{-1}$~Mpc$^{-1}$}  are the
largest  relatively   isolated    density enhancements
in the Universe.   As  commonly
accepted,  superclusters are formed    by  perturbations of the   same
wavelength in the initial density  field (Einasto et al.  \cite{e2001}
and references therein).

Superclusters  have  mainly  been  studied using  the   data on rich
clusters  of   galaxies
(Einasto et al.   \cite{e2001}, and references therein). 
Properties of superclusters  (their shapes and orientations)  have
been   studied     by   West \cite{w89},    Plionis  et  al.
 \cite{pli92},  Jaaniste   et  al.  \cite{ja98} and  Kolokotronis 
et  al. \cite{kol02}.
Already early studies of the fine structure  of nearby superclusters
and  of the  distribution of matter in low density  regions between 
superclusters (Lindner et al. \cite{lind95} and references therein) 
showed  that  superclusters   have  a complicated  structure, 
where clusters and  groups of galaxies are connected
by filaments of galaxies. Superclusters may also contain hot gas 
(Kull \& B\"ohringer
\cite{kull99}, Bardelli et al. \cite{bar00}, Rines et al.
\cite{rin01}, Rose et al. \cite{ros02}).

At present several deep galaxy surveys  are publicly available.  Among
these surveys are the ESO Slice Project survey (ESP, Vettolani et al.
\cite{vet97}), the Las Campanas Redshift Survey (LCRS; Shectman et al.
\cite{she:she}),   the 2 degree  Field  Galaxy  Redshift Survey  (2dF,
Colless et al.  \cite{col01}) and  the Sloan Digital Sky Survey (SDSS,
York  et al. \cite{yo0}).  These  surveys   can be  used to study  the
structure of  a  large number of  superclusters in  more detail and on
larger scales than hitherto possible.

The  catalogue of  loose  groups of   galaxies  extracted from the  LCRS
(LCLGs, Tucker et al.  \cite{tuc:tuc}, hereafter TUC) gives us 
 an   opportunity to study  the  spatial distribution and intrinsic 
properties of  loose groups on large  scales, up to redshifts
$z \approx 0.15$.  In the LCRS, galaxies have  been observed in 6 thin slices;
thus, in order to  use this survey to study the 3D structure of
the Universe, it is necessary to analyse this survey together with data on 
rich clusters of galaxies, e.g. Abell clusters.
Such a combined analysis of the spatial distribution of loose groups, 
Abell  clusters, and superclusters  of Abell  clusters  enables  us to
exploit the  deep     slices to study the    fine  structure  of
superclusters and the hierarchy of the structures in the Universe.

In the present paper we found  populations of  LCLGs in superclusters of
Abell  clusters, using  density  fields and the friends-of-friends  (FoF)
analysis of the LCRS.  We studied the properties of these systems
and  the distribution of LCLGs   with respect to the supercluster-void
network,  traced by Abell  clusters.    We  also studied  the
properties of  LCLGs in systems that contain  no Abell clusters.

In the next  Section we describe our  LCLG  and Abell  cluster 
samples.  In Sect.   3 we identify LCLGs that belong to 
superclusters.  Then we study the properties of superclusters and 
the distribution  of   LCLGs with   respect to  the  supercluster-void 
network. In the last two Sections we give a discussion and summary of 
our results.

\section{Observational data}

\subsection{LCRS loose groups}

The LCRS  (Shectman et al.  \cite{she:she} ) is an  optically selected
galaxy redshift  survey   that extends to  a redshift   of  0.2 and is
composed of six slices, each covering an area of roughly $1.5^{\circ}
\times 80^{\circ}$.  Three of these slices are located in the Northern
Galactic   Cap and are    centred at the declinations $\delta=-3^{\circ},
-6^{\circ}, -12^{\circ}$;  the other three  slices are located  in the
Southern   Galactic  Cap  and are   centred  at the declinations $\delta=
-39^{\circ}, -42^{\circ},  -45^{\circ}$.   The thickness of  slices is
approximately   $7.5$~\Mpc\ at    the   survey's   median    redshift.
Altogether, the  LCRS contains redshifts  for $23,697$ galaxies within
its official photometric and geometric boundaries.

The survey  spectroscopy was  carried out using a 50 fibre multiobject
spectrograph with the nominal apparent magnitude limits
for the spectroscopic    fields $16  \le R   \le  17.3$, and 
 a 112 fibre multiobject
spectrograph with a larger range  of
apparent  magnitudes   ($15  \le  R  \le  17.7$).   
Therefore, the  selection criteria varied  from field to
field, often within a given slice.

Using the FoF percolation  algorithm, TUC extracted the  LCLG catalogue
from the LCRS.  The linking lengths were chosen so that each
group is contained within  a galaxy number density enhancement contour
of $\delta n/n = 80$. When extracting these LCLGs, great care was taken
to account for 
the   radial selection function,  the
field-to-field  selection effects inherent  in the LCRS,
and the boundary effects due to the fact that the LCRS is
composed of six thin slices.
As the  derived properties of the  LCLGs in the 50-fibre
fields do not differ substantially from  the derived properties of the
LCLGs in the 112-fibre fields, the selection effects were successfully 
eliminated. 

The LCLG catalogue contains 1495 groups in the redshift range of $10,000
\le cz \le 45,000$~{\rm km s}$^{-1}$.  This  is one of the first deep,
wide samples of loose groups;  as such,  it enables  us for the  first
time to investigate the spatial distribution and properties of groups in
a large volume.

\subsection{Abell clusters and superclusters} 
We use Andernach \& 
Tago's (\cite{at98}) recent compilation of data on rich clusters of 
galaxies (Abell \cite{abell}  and   Abell et  al. 
\cite{aco}).  From  this compilation we selected  
clusters of all richness  classes out to the redshift $z=0.13$; 
we, however, excluded clusters from the  generally poorer supplementary 
(or ``S'') catalogue of clusters compiled by  Abell et al.   
(\cite{aco}).  Our final Abell cluster sample   contains 1665 clusters,  
of  which 1071 have measured redshifts for  two  or more  galaxies.  
Using  this  sample of   Abell clusters    Einasto et   al.   
(\cite{e2001}) compiled    a catalogue of superclusters of Abell  
clusters (``Abell superclusters'') and  a list of X-ray  clusters in 
superclusters. In the  present paper we use this catalogue as  a 
reference  to study  supercluster  membership of  LCLGs. The X-ray data   
were  also  taken  from  the   lists  by Schwope   et al. 
(\cite{sch:sch},     hereafter    RBS)     and by    de     Grandi  et 
al. (\cite{deg:deg}).  Information on radio clusters has been taken from 
Ledlow \& Owen (\cite{lo95}).

\section{Las Campanas Loose Groups and superclusters} 

Einasto et al.   (\cite{e2001})  identified  
superclusters of Abell  clusters 
(``Abell superclusters'') using the FoF algorithm with 
linking the length $R =  24$\Mpc\, that corresponds approximately to  an 
overdensity contour  of 2.  This radius  was  chosen by
studying the spatial  distribution  of rich clusters  of galaxies. This 
list helps us to identify Abell  superclusters that cross the LCRS slices. 
Using the same linking length  to search for members of  superclusters 
among the loose groups  links almost $90 -  95$\% of groups 
(Fig.~\ref{fig:mflgl2}). At the other extreme, by simply identifying as 
supercluster  members those loose  groups that are 
located around rich clusters 
in a sphere of  radius $R = 6$\Mpc, as done by 
Einasto  et al.  (\cite{env}),  we 
miss   the     outer members   of   superclusters. Thus we have used 
additional  methods to determine the system membership.   
We identified systems of loose groups and then 
searched for those systems that belong to Abell superclusters as 
described in the following two subsections.

\subsection{The FoF approach}

\begin{figure}
\vspace{7cm}
\caption{Multiplicity function of LCLGs showing the fraction of groups
in systems  with  at least  2 members  as a  function of the neighbourhood
radius. Bold  lines: solid line  -- slice $\delta= -3^{\circ}$, dotted
line  -- slice $\delta= -6^{\circ}$,   dashed  line -- slice  $\delta=
-12^{\circ}$.  Thin lines:  solid  line: slice $\delta=  -39^{\circ}$,
dotted line  -- slice $\delta= -42^{\circ}$, and  dashed line -- slice
$\delta= -45^{\circ}$.  Note the difference between the slice $\delta=
-6^{\circ}$ and other slices.  }
\includegraphics{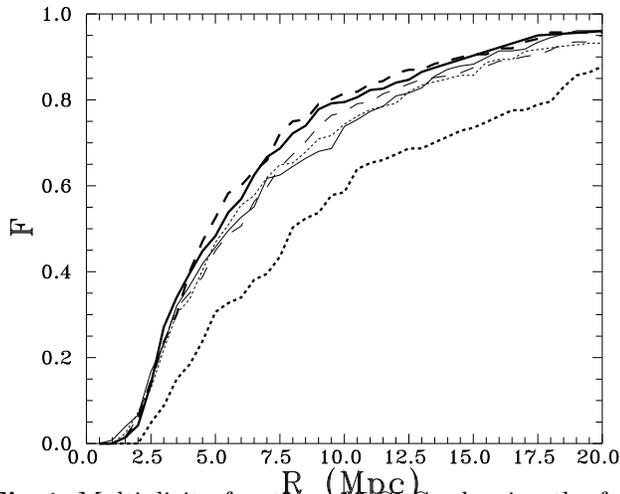}
\label{fig:mflgl2}
\end{figure}

To identify  systems of loose  groups we applied the FoF 
algorithm to LCLGs  from the each Las Campanas slice separately 
using a wide range of linking lengths     
(or   neighbourhood radii). Fig.~\ref{fig:mflgl2} shows the fraction 
of LCLGs in systems with at  least  2 member groups   as a  function of 
the  neighbourhood radius.     This figure shows that at a 
neighbourhood  radius $R = 6$~\Mpc, 50\% -   60\% of groups  belong  to 
such systems.  We chose this radius, $R =  6$\Mpc, as our linking  
length to identify  systems of LCLGs.  The  same   radius  
was used to search for populations of  loose groups around  rich  
clusters in  Einasto et al. (\cite{env}).  A more  detailed analysis of 
the multiplicity functions shows that at this  neighbourhood 
radius richer systems (with at least  8 member groups)  
start to form.  This was one of the reasons 
to use the neighbourhood radius $R = 6$~\Mpc.

Fig.~\ref{fig:mflgl2} shows that the multiplicity function of groups
from the $\delta=   -6^{\circ}$ slice differs from  the multiplicity
functions for groups in  other slices. This may be
due to the  selection   effects, as  in  the  case  of this  slice
all-but-two fields were observed  with the 50-fibre spectrograph, and
the selection   effects that may   decrease the number  of  groups are
stronger.    Another possibility  is  that  this  slice crosses a region
dominated by voids and thus the number of groups in this slice and the
number of superclusters crossed  by this slice   is smaller (see  also
Table~\ref{tab:LGSC}).  To check the first possibility we recalculated
the multiplicity functions, using the neighbourhood radius $R$ in units of
the dimensionless radius,  $r=R/R_0$, where $R_0=[3V/(4\pi N)]^{1/3}$  is
the Poisson radius  (the radius of a sphere  which  contains one particle),
$N$ is  the number  of particles in  the  sample, and $V$ is  the
volume of the sample. The multiplicity function of the slice $\delta=
-6^{\circ}$ still   differed from the  multiplicity functions  of other
slices. Thus this  difference may be at least partly
caused by peculiarities of the
large scale distribution of groups in this  slice.

Using the FoF  technique, we occasionally  find that  some LCLGs, which
appear to be associated  with  an Abell supercluster remain   isolated
even at the linking length $R = 6$\Mpc.   Thus, to link these quasi
isolated LCLGs to the supercluster, we could 
 use either a variable linking length which would link these LCLGs to
the supercluster system, or we could use another approach to determine
system membership.

\subsection{The density field approach}

As another method to identify systems of loose groups and to determine
membership   of  LCLGs within Abell superclusters    we  shall use the
density field of the Las  Campanas Survey slices calculated from the LCRS
galaxy distribution.  Details  of these calculations will be published
elsewhere (Einasto,  J. et al. \cite{ee2003}).  
Here we shall only briefly outline the method.

To calculate  the density field  we formed  a grid of cell  size
1~\Mpc\  and used Gaussian smoothing (Einasto et al. (\cite{je2003}). 
The  thickness of the LCRS slices is
only $1.5^{\circ}$; thus   we calculated 2-dimensional  density
fields. To take  into account the  thickness of the slice, the smoothed
density field was divided by the thickness of the slice at the location
of a particular cell in real 3-D space. 
In this way the  map of the density field
is reduced to that of a planar sheet of constant thickness.

To identify superclusters of galaxies we used the smoothing length
$\sigma_{{\rm sm}}=10$~\Mpc. Numerical simulations have
shown that    this  smoothing length     is suitable   for   selecting
supercluster-size density enhancements (Frisch  et al. 1995; see  also
Basilakos,  Plionis  \& Rowan-Robinson \cite{bpr01}). 

The number of  superclusters has a maximum for
all slices at the relative  threshold density $\Delta_0=1.5$
(Einasto et al. \cite{je2003}). The relative density  $\Delta$ 
is expressed in units of the mean density ($\Delta 
\equiv \rho / \langle \rho \rangle$), averaged over the whole observed 
area covered by a particular slice. For
lower $\Delta_0$ superclusters merge; for higher $\Delta_0$
 fewer high-density regions are counted. 
Our analysis shows that superclusters  are still separated 
at limiting    densities  around
$\Delta_0=  2.0$.  This  value,  $\Delta_0= 2.0$,  defines compact and
rather rich  superclusters. 

The    density contrast   in the large    scale   environment of  
superclusters (in  the regions around Abell  clusters that belong to 
superclusters)  varies    around $\Delta_0= 2.0$. Therefore, in 
order to identify  populations of
loose groups that belong to  superclusters, we use 
a variable threshold density limit
to determine superclusters.
In addition, we use the  value of 
the overdensity  in  superclusters as one  of the quantitative 
characteristics of the systems (Table~\ref{tab:LGSC}).

In   Table~\ref{tab:LGSC}   we  list 
superclusters, which intersect LCRS  slices. 
In total we find 19 systems, 16 of which belong to Abell 
superclusters and 3 of which are relatively  isolated.    We  denote the    
sample  of  loose groups   in  Abell superclusters as   LCLG.scl,  and  
the  sample of   loose   groups  in high-density systems without  Abell  
clusters (LG superclusters) as LCLG.lgs.

\begin{table*}
\begin{center}
\caption{The data about loose groups in superclusters of Abell clusters 
and in rich systems without Abell clusters}
\tabcolsep 5pt   
\begin{tabular}{rrrrrrrrrrr}
\hline
$N_{\rm SCL}$ & R.A. & D & $N_{\rm Abell}$ &  $N_{\rm ALC}$ &$N_{\rm X}$ & $N_{\rm LCLG}$ 
& $\Delta$& a & b/a & (ar) \\
(1)& (2) & (3) & (4) & (5)& (6) & (7) & (8) & (9) & (10) & (11) \\
\hline
\multispan3 Slice $\delta=-3^{\circ}$  &&& &&&&&\\
\hline              
   88  & 10.3 & 175 & 5 & - &   & 23  & 3.5  & 22.9 & 0.49 & 79 \\
   98  & 11.5 & 180 & 3 & 1 &   & 6   & 1.3  & 11.1 & 0.18 &  84 \\
  100  & 11.3 & 280 & 9 & 3 &   & 23  & 2.6  & 68.7 & 0.45 &  12 \\
  126  & 12.9 & 240 & 7 & 4 & 3 &  7  & 5.8 & 26.6  & 0.32 &  88 \\
  155  & 15.2 & 280 & 2 & 1 &   & 3   & 1.5  &      & &     \\ 
&&&&&&&&\\
 slg1  & 14.6 & 170 &  &   & & 8   & 2.2   & 10.2  &  0.32 & 67 \\
                                                
\hline                                            
\multispan3 Slice $\delta=-6^{\circ}$  && &&&& \\
\hline               
   88  & 10.1 & 160 & 5 & 1 &   & 6   & 4.4 &  9.2 & 0.29  & 64 \\
  100  & 11.3 & 281 & 9 & 1 &   & 3   & 1.6 &  && \\
  267  & 11.9 & 348 & 4 & 1 &   & 4   & 4.1 &  && \\
\hline               
\multispan3 Slice $\delta=-12^{\circ}$  && &&&& \\ 
\hline               
   88  & 10.2 & 169 & 5 & 1 & 1 & 2   & 0.9 &  & & \\
  119  & 12.6 & 272 & 6 & 1 & 1 & 15  & 4.3 & 21.0 & 0.20 & 87 \\
  141  & 13.9 & 198 & 4 & 1 & 1 &  7  & 2.7 & 12.9 & 0.38 & 62 \\
  156  & 15.1 & 308 & 2 & 1 &   & 10  & 2.5 & 27.1 & 0.36 & 40 \\
  274  & 13.8 & 248 & 2 & 1 &   & 3   & 3.0 &  & & \\
&&&&&&&&\\
 slg2  & 11.6 & 219 &  &   & & 8   & 3.3   & 9.3 & 0.86 & 31 \\   
\hline
\multispan3 Slice $\delta=-39^{\circ}$  && &&&&\\
\hline
    5  &  0.3 & 328 &  5 & 2 & &  4   & 2.2  & & &\\
    8  &  0.6 & 182 &  3 & 1 & &  2   & 2.6  & & &\\
    9  & 23.5 & 275 & 25 & 4 & & 14   & 2.4  & 49.6 & 0.41 & 53\\
   22  &  0.9 & 335 &  6 & 1 & &  2   & 1.0  &  & &\\
   23  &  1.1 & 216 &  2 & 2 & & 11   & 1.8  & 26.4 & 0.33 & 72 \\
   37  &  2.0 & 287 &  5 & 1 & &  1   & 1.6  &  & & \\
   48  &  3.4 & 180 & 35 & 4 & & 24   & 4.2  & 47.5 & 0.58 & 21\\
  220  & 23.9 & 144 &  3 & 1 & &  3   & 1.7  &  && \\

\hline
\multispan3 Slice $\delta=-42^{\circ}$ &&&&&& \\
\hline
    9  & 23.3 & 252 & 25 & 1 & &  2   & 3.7  & &  &\\
   37  &  2.0 & 289 &  5 & 2 & &  2   & 1.2  & &  &\\
   48  &  3.4 & 180 & 35 & 1 & &  9   & 4.8  &20.4 & 0.29 & 55 \\
  182  & 21.5 & 205 &  6 & 4 & & 18   & 2.5  &34.0 & 0.37 & 30 \\
  222  &  0.2 & 255 &  2 & 2 & &  4   & 3.2  & & &\\
\hline
\multispan3 Slice $\delta=-45^{\circ}$ &&&&&&\\
\hline
   48  &  3.4 & 190 & 35 & 4 &2& 14   & 4.3  & 25.5 & 0.47 & 63 \\
  182  & 21.6 & 188 &  6 & 1 & &  5   & 2.1  &  & &  \\
  183  & 21.2 & 270 &  2 & 2 & &  9   & 2.5  & 25.1 & 0.29 & 2 \\
  197  & 22.6 & 252 & 11 & 2 & &  6   & 2.3  & 14.4 & 0.26 & 64 \\
  206  & 22.9 & 342 &  4 & 1 & &  4   & 3.6  &  & &  \\
&&&&&&&\\
 slg3  &  1.3 &  250 & &   & & 8   & 3.0   & 9.35 & 0.67 & 79\\

\hline
  
\label{tab:LGSC}
\end{tabular}
\end{center}
{\tiny 
The columns in the Table are as follows:

\noindent Column (1):
The supercluster number from the 
catalogue of superclusters of Abell clusters by Einasto et al. 
(\cite{e2001}). 

\noindent Columns (2): The R. A. of the supercluster (hours).

\noindent Columns (3): The distance to the supercluster (in \Mpc).

\noindent Columns (4): The number of Abell clusters in the supercluster.

\noindent Column (5): The number of Abell clusters in the part of 
supercluster inside the LCRS slice.

\noindent Column (6): The number of X-ray clusters
among Abell clusters in this part of the supercluster
that is inside LCRS slice.

\noindent Column (7): The number of LCLGs in the supercluster region. 

\noindent Column (8): The maximum relative density in the supercluster
region, $\Delta$, in units of the mean density.

\noindent Column (9): The length of the largest semiaxis $a$ of the 
ellipsoid of concentration of LCLG superclusters 
(in \Mpc).

\noindent Column (10): The ratio  of semiaxis $b/a$ of  the ellipsoid
			of concentration.

\noindent Column (11): The angles between line-of-sight 
and the large semiaxis $a$ of LCLG superclusters (in degrees).
}
\end{table*}

The colour figures and the three-dimensional distribution of LCLGs, 
rich   clusters and superclusters can be seen at the home page of 
Tartu Observatory ({\tt http://www.aai.ee/$\sim$maret/cosmoweb.html})
and via EDP ({\tt http://www.edpsciences.org}).

\section{Systems of Las Campanas Loose Groups}
\subsection{Approximation by the ellipsoid of concentration}

Superclusters (Einasto et al. \cite{e2001}) are not
regular systems with  well-defined boundaries but aggregates  of quite
sparsely distributed  clusters, groups and galaxies
  with  some central  concentration.  To
find the boundaries of such superclusters and to study their shape and
orientation  we  approximate   the   spatial distribution   of   objects
(clusters, groups,  galaxies) in superclusters
by a 3-dimensional ellipsoid  of concentration. For such an  ellipsoid
we can find  the centre, volume and principal  axes.  Although in most
cases our superclusters do not form a regular body, these parameters
help us to  describe  the density and  alignments  of the  elements  of
large-scale structure.

In the present study we use the classical mass ellipsoid
(see e.g. Korn \& Korn \cite{korn61}): $$
\sum_{i,j=1}^3\left(\lambda_{ij}\right)^{-1}x_ix_j=5,  \eqno(1)
$$
where
$$
\lambda_{ij}={1\over{N_{cl}}}\sum_{l=1}^{N_{cl}} 
{(x^l_i-\xi_i)(x^l_j-\xi_j)}, \eqno(2) $$
is the inertia tensor for equally weighted groups, $N_{cl}$ is the 
total number of groups, and     $\xi_i={1\over
N_{cl}}\sum_{l=1}^{N_{cl}}x^l_i$ determines the Cartesian coordinates 
of the centre of mass of the system.

The  formula determines a  3-dimensional  ellipsoidal surface with the
distance from the centre of  the ellipsoid equal  to the rms deviation
of individual objects in the corresponding direction.  This method can
be applied  for superclusters with $N\ge 5$. The problems related to the
stability of the method and the influence  of observational errors have
been discussed in Jaaniste et al. (\cite{ja98}).

In the case   of  contemporary deep  surveys as  the  LCRS, where  the
galaxies with measured redshifts cover only  a narrow slice, the shape
of the ellipsoid strongly depends on the parameters  of the survey.  For a
thin layer  it  is possible to  use  a 2D approximation,  ignoring the
third coordinate (in our case, the declination).  Since the 
 geometry  of slices   is far from a
plane-parallel sheet we shall use the 3D algorithm to approximate the
shapes  of systems. In this  way we get 3D  objects that correspond to
the  3D  ``slices''  cut    from  the larger  3D   superclusters  (Abell
superclusters).

\begin{figure}
\vspace{7cm}
\caption{The angle between the line-of-sight and the large semiaxes of 
the LCLG superclusters (Table~1). The diagonal corresponds to the uniform 
distribution. 
}
\includegraphics{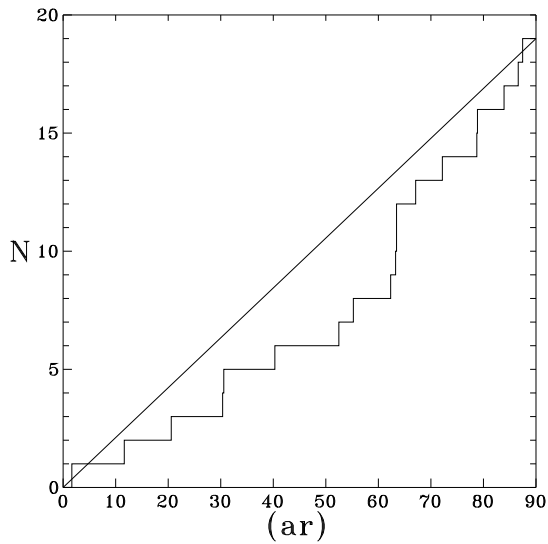}
\label{fig:lor}
\end{figure}

In Table~\ref{tab:LGSC} we present some parameters of the ellipsoid of
concentration for all systems composed of  at least 6 LCLGs.  
In Fig.~\ref{fig:lor} we plot the distribution of angles between the
line-of-sight   and  the large      semiaxes  of the LCLG   superclusters
(Table~1). The diagonal corresponds to a uniform distribution.
The systems are moderately  elongated
(the mean   axes   ratio  3:1)  and   show a   tendency    to  be oriented
perpendicularly  to  the line  of sight.   The  same tendency has been
found for Abell superclusters (Jaaniste et al. \cite{ja98}).

\subsection{LCLGs and Abell superclusters}

Here we  describe shortly   the  most prominent  Abell  superclusters
crossed by the LCRS slices.    

\begin{figure}
\vspace{8cm}
\caption{The distribution of Abell clusters
(open circles), Las Campanas Loose
Groups (trialngles) and LCRS galaxies (stars)
in the region of the supercluster SCL126
along line-of-sight, with shape ellipses. 
Open circles denote Abell clusters, 
triangles -- LCLGs, stars -- LCRS galaxies. 
}
\includegraphics{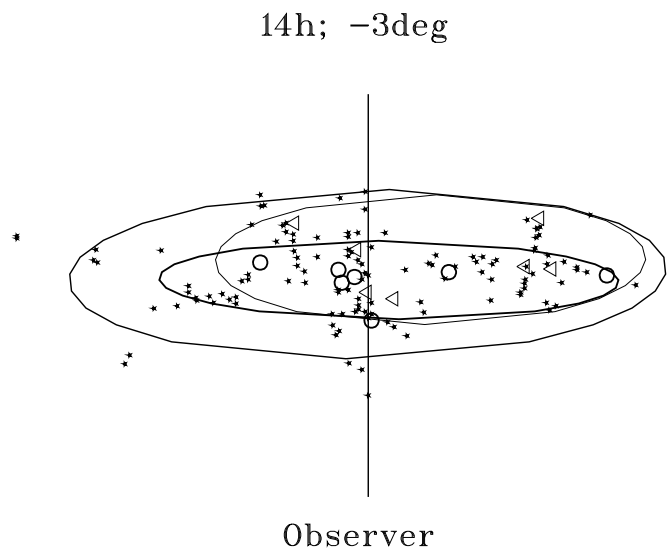}
\label{fig:s126e}
\end{figure}

The most prominent Abell  supercluster
crossed by the Northern LCRS slices is  the supercluster SCL126 in the
direction of the Virgo constellation (Fig.~\ref{fig:s126e}).
Four Abell clusters of total seven member clusters of this supercluster
are located in the  Las Campanas slice  $\delta= -3^{\circ}$  within a
sphere of a diameter  of about  $10$~\Mpc.  Three   of  these four
clusters are strong  X-ray sources.  The fifth  X-ray  cluster in this
supercluster is Abell  1750, but it is located  outside the  slice.
This cluster is a merging binary cluster  (Donelly et al. \cite{don}).
Four Abell clusters  in this supercluster are radio sources.  Such a
concentration  of  rich optical,  X-ray, and    radio clusters in  one
supercluster in a very small volume makes SCL126 one of the most
unusual superclusters  currently   known.  
There are three  LCLGs in the   central area of the  supercluster. All
these groups are unusually rich.
Table~\ref{tab:LGSC} shows that the local density  in the area of this
supercluster is the largest in the whole survey,  $\Delta > 5$.

Another  rich supercluster in  this slice  is the supercluster SCL100 
with 9 member Abell  clusters (the supercluster Leo A). The members of this  
supercluster  are very  close to  the LCRS  slices being located  almost  
all in one  plane.  Three  member  clusters lie in the slice    $\delta= 
-3^{\circ}$,  and  one  in the slice  $\delta= -6^{\circ}$. Altogether 
there are 26 loose groups from the LCRS in this supercluster. However, 
the properties  of this supercluster  differ from those  of   SCL126.  
There  are  no  X-ray clusters  among  these clusters.  One cluster, 
Abell 1200,  is a  radio source. The distances between the Abell clusters in 
this  supercluster are quite large, and this supercluster   resembles      
rather   a   filament      of  clusters. Table~\ref{tab:LGSC}  shows  
that this  filament-like  supercluster is located almost along the 
line-of-sight.

A part of    the Sextans supercluster  (SCL   88) is seen   in all three
Northern slices. The Abell  cluster members  of  the Sextans supercluster
closest to the LCRS slices are A978 (the slice $\delta = -6^{\circ}$), and
A970 (the slice $\delta= -12^{\circ}$, this is an X-ray cluster). In the slice
$\delta= -3^{\circ}$ 23 groups  form a system that  is an extension of
this supercluster. The groups themselves  in this extension are relatively
poor -- the   richest loose group  here has  $N_{\rm ACO} =   29$, being
poorer than Abell clusters of   richness class $R   = 0$. 

Another supercluster, seen in  the slice $\delta= -12^{\circ}$, is the
supercluster SCL119. Its member cluster  Abell 1606 is an X-ray
source and is associated with three loose groups.

\begin{figure}
\vspace{15cm}
\caption{The distribution of Abell clusters and Las Campanas loose
groups    in   the   region     of  the    supercluster  SCL48    (the
Horologium-Reticulum supercluster) in  supergalactic YX  (upper panel)
and YZ (lower panel) coordinates  in Mpc. Filled circles denote  Abell
clusters, open circles  --  LCLGs from the slice  $\delta=  -39^{\circ}$),
crosses --  LCLGs from the slice $\delta=  -42^{\circ}$), and x's -- LCLGs
from the slice  $\delta= -45^{\circ}$).  The Abell  numbers of the richest
Abell clusters in this supercluster are given (see text).  }
\includegraphics{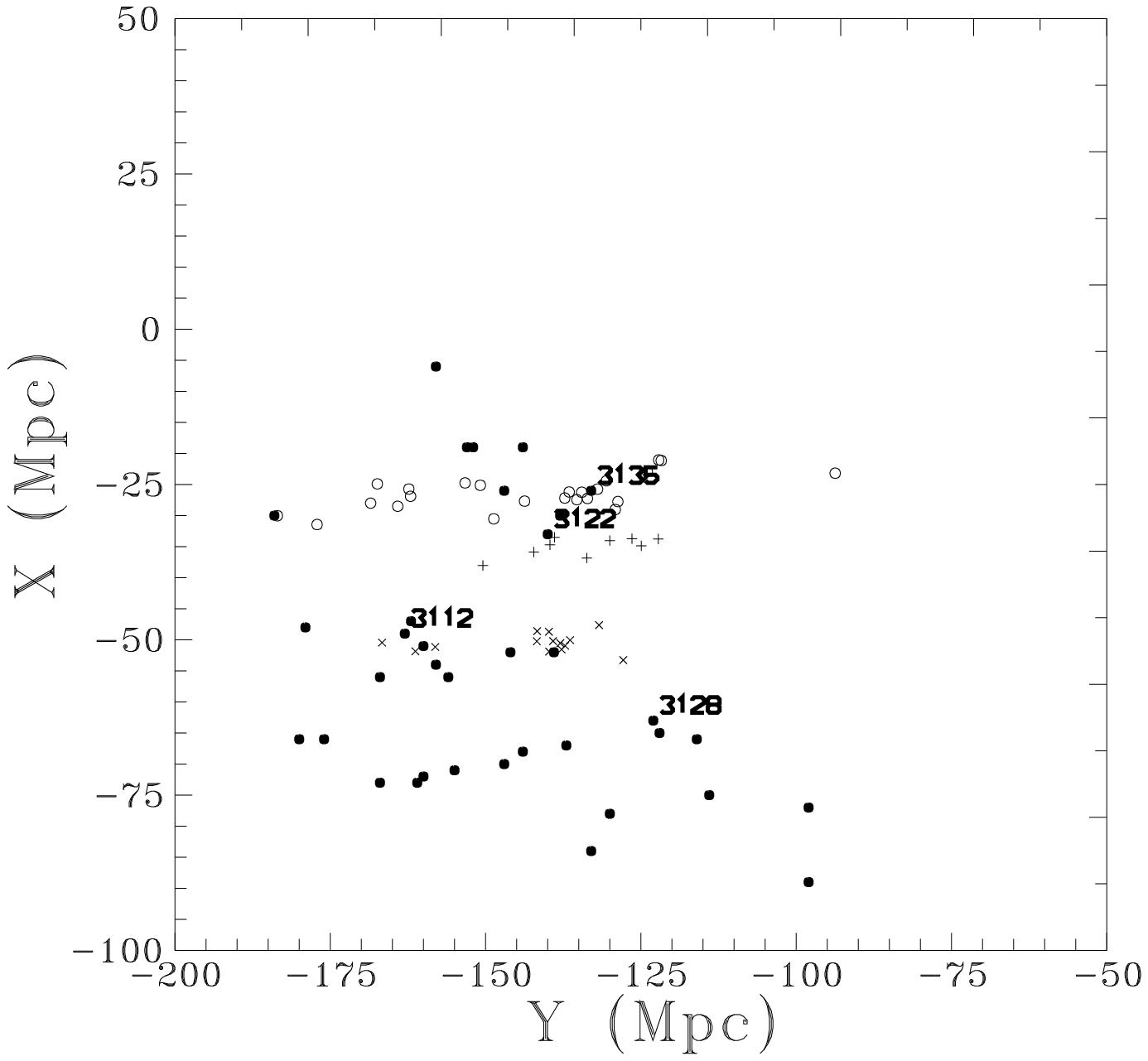}
\includegraphics{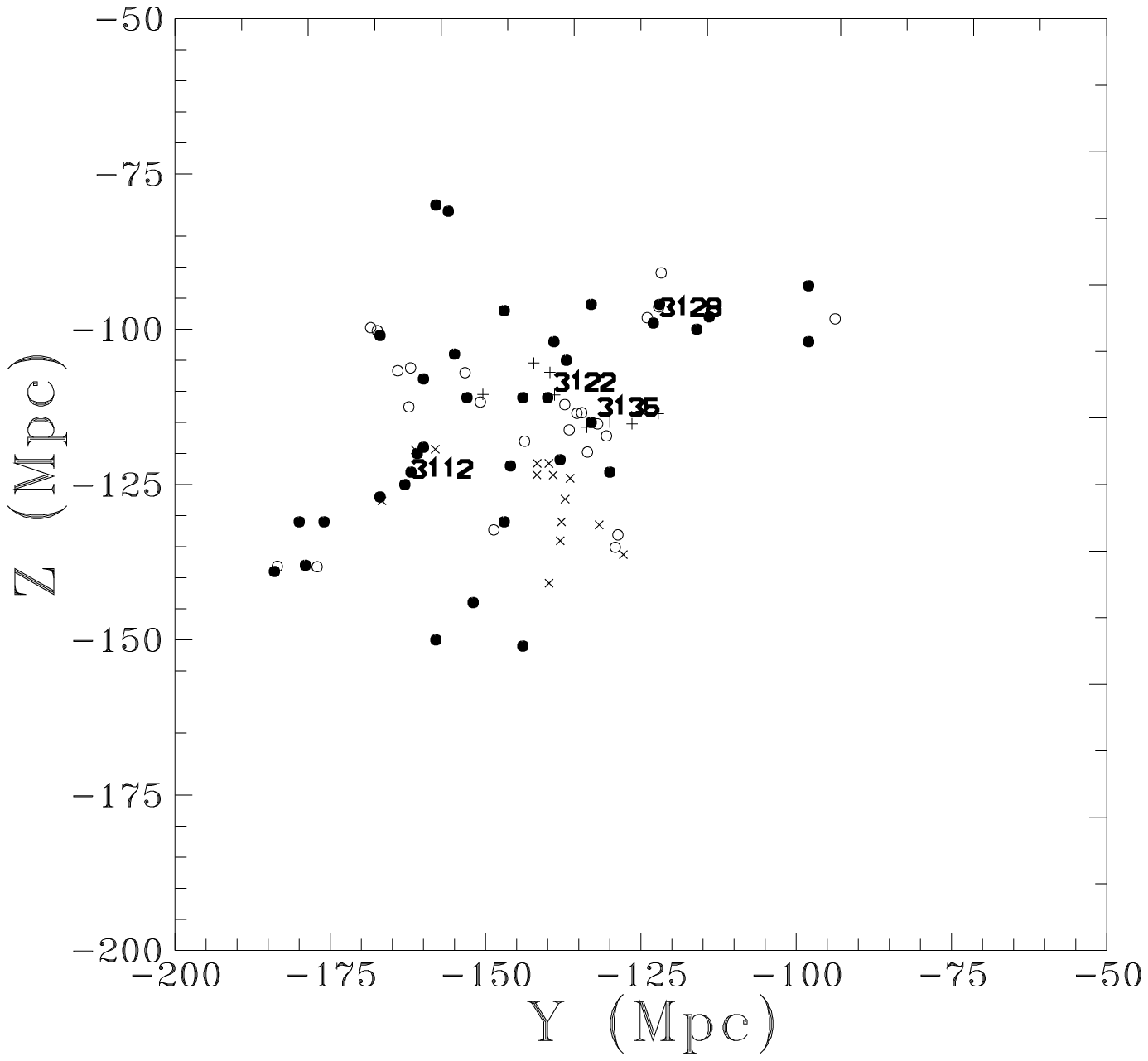}
\label{fig:hryx}
\end{figure}

The  most prominent supercluster  crossed  by all Southern LCRS slices 
(and one  of   the richest  superclusters known) is    the Horologium - 
Reticulum supercluster  (SCL48), 47  LCLGs being associated with this 
supercluster  (Fig.~\ref{fig:hryx}). This  supercluster   contains two  
X-ray clusters  and a number  of  APM clusters (Einasto et al. 
\cite{e2002}). One concentration  of Abell clusters and LCLGs in this  
supercluster is centred  on the very rich Abell cluster A3135  
(the richness class $R = 2$,  the slice $\delta=  -39^{\circ}$) that is 
associated with 7 loose groups. Another concentration of clusters in the   
Horologium-Reticulum  supercluster  is   crossed by the  slice $\delta=  
-45^{\circ}$. The richest  Abell cluster in this region is Abell 3112,  
an X-ray and radio   source. However,  the richest concentration 
of LCLGs in this slice  is located around another member cluster of this 
supercluster, Abell 3133 (the richness class $R = 0$). A third concentration  
of groups and clusters  in this supercluster is located  around the Abell 
cluster 3128  (Rose  et al. \cite{ros02}) at the distance   of about   
$40$\Mpc\ from the   cluster  Abell  3135.  This concentration,   
however, lies outside  the   boundaries  of the  LCRS slices.  All these  
concentrations    are connected by  filaments   of galaxies, groups and   
clusters that surround underdense regions (see also Rose et al. \cite{ros02}).

Another very  rich supercluster  crossed  by  the  LCRS slices is  the
Sculptor  supercluster (SCL9). There  are   five Abell clusters  in the
region of the LCRS slices from  this supercluster, 4 in the slice $\delta=
-39^{\circ}$ and  1 in the slice  $\delta= -42^{\circ}$.  Altogether there
are 16 loose groups near these rich clusters in this supercluster.

The supercluster SCL23 in the slice  $\delta= -39^{\circ}$ consists of
two Abell  clusters, Abell  2860  (a radio source)  and Abell
2911. There are 11 LCLGs in this region.  This supercluster is seen in
the ESP survey  as  a very strong  density enhancement  in  the galaxy
distribution (Vettolani   et  al.  \cite{vet97}).   This  supercluster
separates two voids, each of which have diameters of about $100$ \Mpc.
On opposite sides of these voids are  the Horologium-Reticulum and the
Sculptor superclusters.

The Southern slices   $\delta= -42^{\circ}$ and  $\delta= -45^{\circ}$
cross  the supercluster SCL182.  All 6 member Abell clusters
of this supercluster are located in these slices. The cluster A3809 is
an X-ray source.

\subsection{Rich systems of loose groups only}
Rich systems (superclusters) of  LCLGs
(Table~\ref{tab:LGSC}) contain at least 8 LCLGs for the linking length 6
\Mpc\, and have no Abell clusters.  

One system consisting of 8 
loose groups is located in the  slice $\delta= -3^{\circ}$ in the void
separating the Abell  superclusters  SCL 88 (Sextans),  126,  and 155. 

Another such   high-density system of loose groups   is located in the
slice $\delta= -12^{\circ}$  at   a distance of about  $220$\Mpc.  The
closest Abell  cluster  to this system  is  Abell  1317, that lies at a
distance of about   15 \Mpc\ from  the  richest group in this  system,
LCLG-12~092. This supercluster has a ``spider-like'' appearance.

In the slice $\delta= -45^{\circ}$ there is a system of 8 loose groups
located between voids at a distance of about $250$\Mpc.  An about $100$
\Mpc\   void     separates  this system    from     the   Horologium-Reticulum
supercluster. There are some galaxy systems in this  void, but no rich
clusters  or  superclusters. This system resembles the Great Wall,
a rich filament   of  galaxies  and  groups   of  galaxies  connecting
superclusters.

The existence of  supercluster systems   that  contain only  LCLGs  and 
separate huge  voids  of diameter of  about  $100$ \Mpc\  agrees with 
the earlier  findings  by Einasto  et  al.   (\cite{e1997})  and Frisch  
et al. (\cite{fri95}). These earlier studies found that huge voids in 
the supercluster-void network are of  similar size, about $100$ \Mpc,  
but the properties of void walls range from those of poor superclusters 
to very rich superclusters containing tens of rich (Abell) clusters.

\section{Properties of LCLGs in different systems}

Let us now compare the properties of LCLGs in Abell superclusters with the
properties  of  LCLGs in rich   systems of LCLGs   containing no Abell
clusters (LCLG.scl and LCLG.slg, respectively; the sample LCLG.slg
includes also the outer members of SCL88 in the slice 
$\delta= -3^{\circ}$). 

Several physical properties have been calculated for each group in the
LCLG  catalogue (TUC).   These  include the  observed number  of group
member galaxies  $N_{\rm obs}$,  the line-of-sight velocity rms
$\sigma_{los}$, the virial  mass  $M_{\rm vir}$, the  total luminosity
$L_{\rm tot}$, and the Abell counts $N_{\rm ACO}$. We refer to TUC for
details of how these properties were estimated.
In Table~\ref{tab:med} we
give the  values of  these properties  for loose groups  from 
different systems, as well as for the total sample of LCLGs.

\begin{table*}

\caption{Median and upper quartile (in parentheses) values of LCLG properties.}

\begin{tabular}{lrrrrrr}
Sample & $N_{\rm group}$ & $N_{\rm obs}$ & $N_{\rm ACO}$
&  $\sigma_{\rm los}$ 
& $\log M_{\rm vir}$ & $\log L_{\rm tot}$  \\
& & &&   km~s$^{-1}$ & $h^{-1}M_{\odot}$  & $h^{-2}L_{\odot}$\\
(1)& (2) & (3) & (4) & (5)& (6) &(7)   \\

\hline 
LCLG.scl  &   204 & 4.5 (8.0)&  15.5 (25.0)& 215 (290)& 13.45 (14.05)&11.15 (11.65) \\
LCLG.slg &  47 & 4.5 (7.5)&  9.5 (15.2) & 175 (280)& 13.25 (13.70)&10.95 (11.20) \\
&&&&&&\\
LCLG.all   & 1495 &  4.0 (5.5)&  16 (25.0) & 164 (270)& 13.20 (13.75)&11.10 (11.45)  \\

\hline
\end{tabular}
\label{tab:med}

The columns are as follows:

\noindent Column (1):
Sample identification (see text).  

\noindent Column (2): 
$N_{\rm group}$, the number of LCLGs in the sample.

\noindent Column (3): 
$N_{\rm obs}$, the observed number of LCRS galaxies in groups.

\noindent Column (4): 
$N_{\rm ACO}$, the group Abell counts.

\noindent Column (5): 
$\sigma_{\rm los}$, the group rms line-of-sight velocities  (in
units of km~s$^{-1}$).

\noindent Column (6): 
$ M_{\rm vir}$, the virial mass of a group  
(in units of $h^{-1}~M_{\sun}$).

\noindent Column (7): 
$L_{\rm tot}$, the total group luminosity in the LCRS $R$-band (in
units of solar luminosity ($h^{-2}~L_{\sun}$)).

\end{table*}

Two  measures  of  a  group's   richness are its  observed  number  of
galaxies,  $N_{\rm obs}$,  and its Abell  count,  $N_{\rm ACO}$,
calculated taking into account the selection
effects  (see  TUC).  We  see  that if we   use these estimated Abell
counts as  a   measure of the  group  richness,  loose groups  in   Abell
superclusters tend to  be richer than loose  groups in systems without
Abell clusters.  Only the two richest  groups in the sample of systems
of  loose groups, LCRS.slg,  have the Abell counts larger  than 30 -- this
population consists of intrinsically  poor loose groups.  In contrast,
in  the population of  loose groups  in Abell  superclusters the  mean
Abell count $N_{\rm ACO} = 20$ and more than 15\% of loose groups have
$N_{\rm  ACO}$ larger than  30.  The richest  group in this sample has
the Abell count $N_{\rm ACO} = 120$, equivalent  to a richness class $R=2$
cluster.

Additionally, let us take as  an example the Abell supercluster SCL222
that  consists  of two Abell clusters    and is  probably completely
embedded within  the LCRS slice  $\delta= -42^{\circ}$.  In  this supercluster
alone there  are 4 loose  groups,  and three of   them are richer than
$N_{\rm ACO} = 30$  -- richer  than  any  group from the
sample of loose groups from systems  without Abell clusters.  This may
indicate that, although the local density  of groups is rather high in
the systems without Abell clusters, all loose  groups in these systems
are  intrinsically  poor. This is  a  hint that  the  absence of Abell
clusters in these systems is due  to the poorness of individual groups
in these region and not due to  possible incompleteness in the Abell
catalogue.

The Kolmogorov-Smirnov test shows that  the  differences  in  the
distribution of group  richnesses between   the  two   samples   are
statistically significant at the 90\% confidence level.

The rms line-of-sight velocity  of loose groups,  $\sigma_{\rm
los}$,  in the  Abell superclusters is  about  $1.2$ times larger than
that of loose groups  from systems without Abell clusters  (LCRS.slg).
There are no loose groups  in this population with the rms velocity 
larger than $\sigma_{\rm los} = 425$~km~s$^{-1}$. The Kolmogorov-Smirnov
test shows that    the differences in  the   distribution of group rms
velocities between  the two group samples  are statistically
significant at the 99\% confidence level.

Comparison of virial masses of  loose groups, $M_{\rm vir}$, shows
that  loose groups in Abell superclusters  have  masses that are about
$1.6$ times   larger than masses of loose   groups in  systems without
Abell clusters. The Kolmogorov-Smirnov   test shows  that  the
differences in the distribution of group virial masses between the two
samples are statistically significant at the 75\% confidence level.

The total luminosities of groups, $L_{\rm  tot}$,  show that  loose
groups in Abell superclusters  are about $1.6$ times more luminous
than   loose   groups   in   systems   without   Abell    clusters. The
Kolmogorov-Smirnov    test shows that    the  differences  in  the
distribution  of group luminosities  between the two samples of groups
are statistically significant at the 95\% confidence level.

\begin{figure}
\vspace{7cm}
\caption{The rms line-of-sight velocities of loose groups in 
Abell superclusters  against  the distances   from the nearest   Abell
cluster in a supercluster. The open circle represents
LCLG-12 163, that is located at an end of a filament centred on
an Abell cluster from SCL119.}
\includegraphics{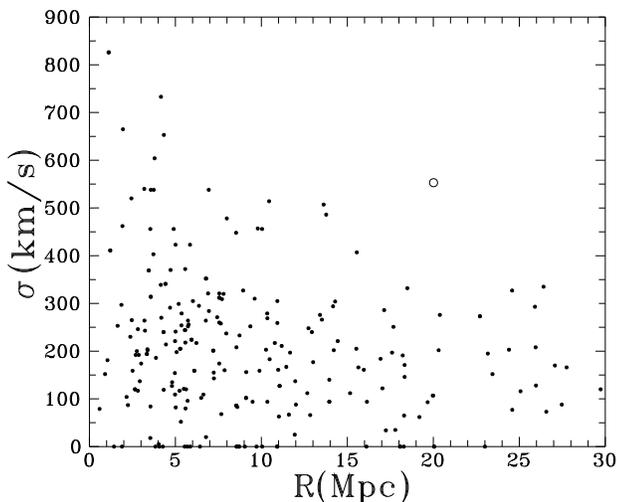}
\label{fig:scd}
\end{figure}

In   order to  study how  far   the environmental
enhancement of loose  group properties extends,  we calculated for loose
groups   in Abell superclusters   the  distance to  the nearest  Abell
cluster.   In  Fig.~\ref{fig:scd}  we plot the rms line-of-sight velocities
of loose groups in  superclusters against the distance to
the nearest  Abell  cluster in   a supercluster. This  figure shows  a
decrease in the velocity dispersions of loose  groups with an increase
in  the distance to the nearest  Abell cluster. Enhancement of 
properties  of loose groups extends quite  far from Abell clusters, up
to about    $15 - 20$~\Mpc.  At   distances  larger than approximately
$20$~\Mpc\  (in our sample  these loose groups  are distant members of
the supercluster SCL 48,  the Horologium-Reticulum supercluster), this
phenomenon becomes weaker.

The loose groups from an outer population of the supercluster SCL88 in
the slice $\delta= -3^{\circ}$ lie  at  distances of about  $16$\Mpc\ from
the nearest Abell cluster in  this supercluster (this Abell cluster is
seen in the LCRS slice $\delta= -6^{\circ}$).
The nearest Abell cluster to the loose groups  in the LCRS.slg1 system
is located at  a distance of  about $30$~\Mpc.   Likewise, the nearest
Abell  clusters to the  loose groups  in  the LCRS.slg2  and LCRS.slg3
systems  are at  distances of   about  $15$~\Mpc\ and about  $35$~\Mpc,
respectively.  Thus,  groups not belonging  to  superclusters or those
groups  which  are    outer members of   superclusters    do not  show
environmentally  enhanced   properties  as do   the  inner  members of
superclusters.

To summarise,  these results extend  the  environmental enhancement of
mass,   velocity dispersion,  and luminosity  of  loose  groups in the
vicinity  of  rich clusters  of  galaxies   found  by Einasto  et  al.
(\cite{env})    to    the   loose    groups within   Abell
superclusters. This effect is absent in systems  which contain no rich
clusters.  We found indications that this effect is also absent in the
case of loose groups from outer parts of superclusters.

\section{LCLGs and the supercluster-void network}

The  large scale distribution of Abell  clusters and superclusters was
described in Einasto   et al. (\cite{e1997}).   In  particular, it was
shown that 75\% of very rich  superclusters are located in the so-called
Dominant Supercluster Plane (DSP) that crosses  the Local Supercluster Plane
at almost right   angles and consists of  chains  of superclusters and
voids between them. Let us now study the distribution of LCLGs with
respect to Abell superclusters and the Dominant Supercluster Plane.

\begin{figure*}
\vspace*{12cm}
\caption{Upper panel: the Abell clusters in equatorial coordinates. Filled
circles show the Abell clusters located in superclusters of richness 8 
and more members, open circles mark the Abell clusters in less rich
superclusters. Solid lines show the location of the LCRS slices, 
dashed lines -- the location of the Dominant Supercluster Plane. 
Lower panels show intersections of supercluster ellipsoids with
the LCRS slices}
\includegraphics{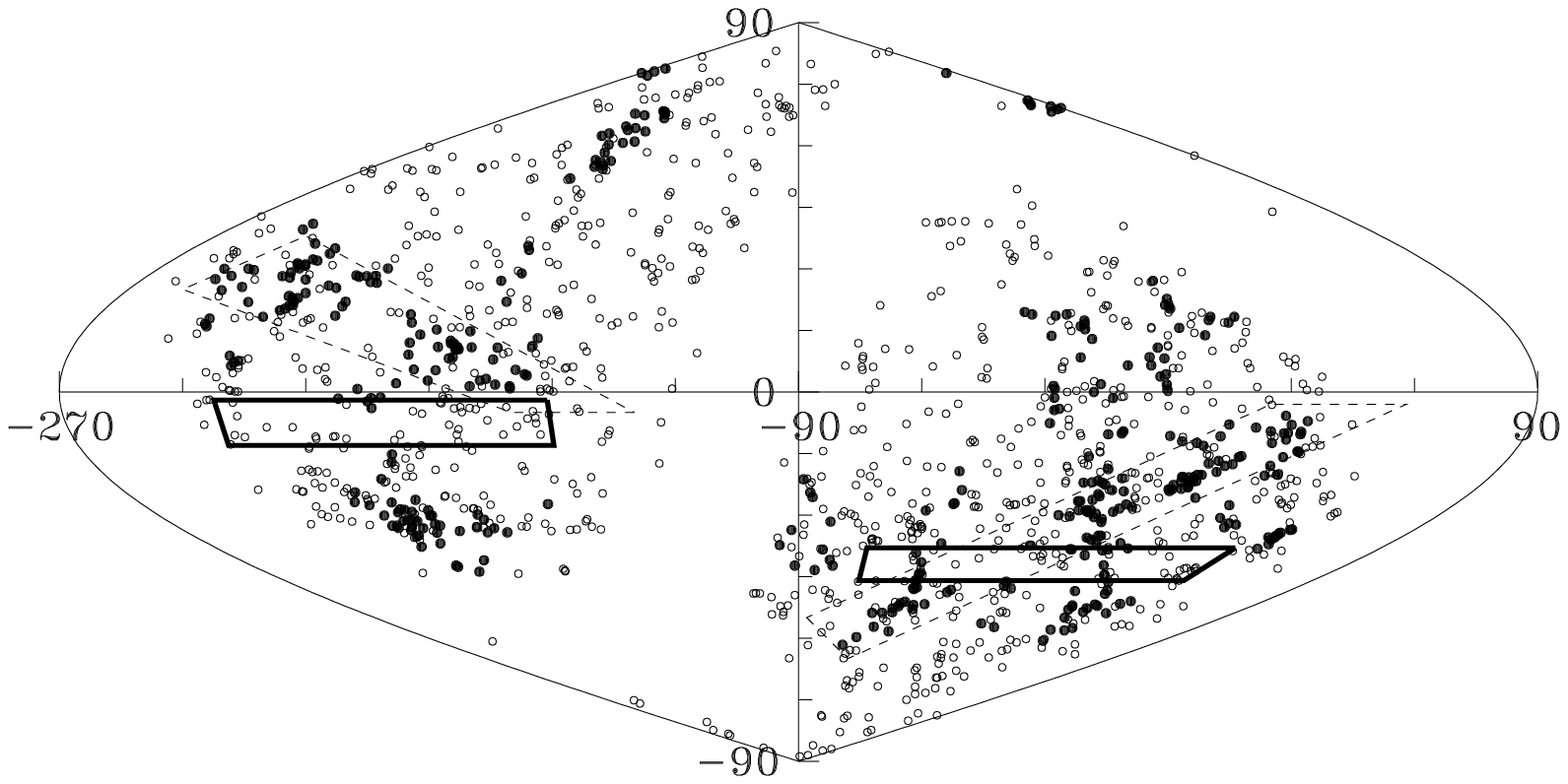}
\includegraphics{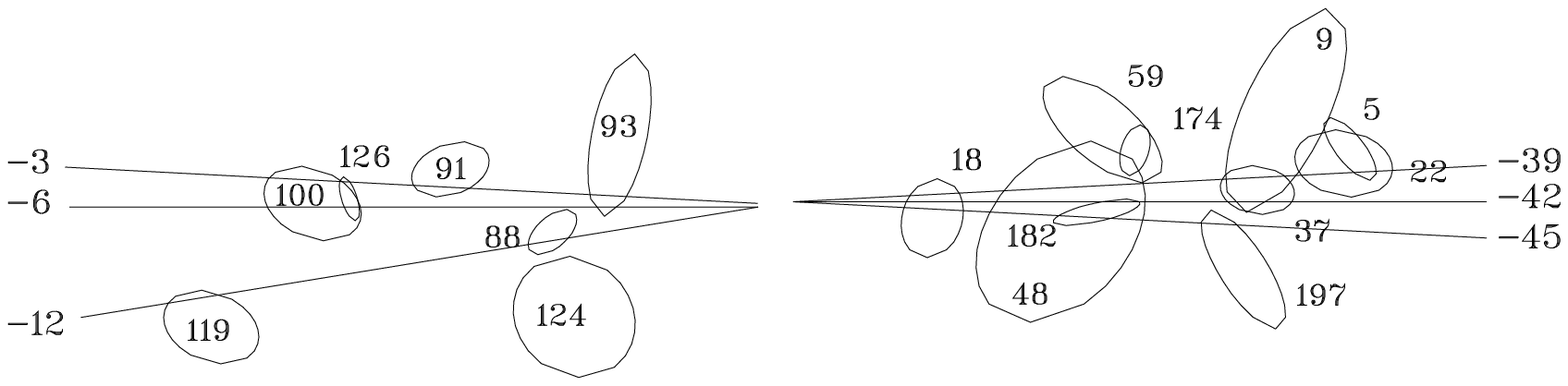}
\label{fig:a48slg}
\end{figure*}

Fig.~\ref{fig:a48slg}  shows the distribution  of  Abell clusters and 
the location  of the LCRS slices  with  respect to  the supercluster-void 
network. The Las Campanas slices cross  several rich superclusters in  the 
Dominant Supercluster Plane:   the     Sculptor     supercluster    
and   the Horologium-Reticulum supercluster  in the Southern  sky, and 
the Leo A supercluster in the Northern sky. The Southern slice at  
$\delta=  -39^{\circ}$ goes almost through  the DSP.  
Of  the Northern slices, the  slice 
at $\delta= -3^{\circ}$ is closest to the  DSP;  the  other Northern 
slices     cross the voids   between superclusters. The Northern 
slice at   $\delta= -6^{\circ}$ crosses the region   most devoid of 
galaxies and galaxy systems. This may  be one of the reasons why the 
number of LCLGs in this slice is  much smaller than the number of groups  
in other slices.

\section{Discussion}

\subsection{Selection effects} 
\begin{figure}
\vspace{7cm}
\caption{Luminosities of LCRS loose groups from superclusters 
of Abell clusters (filled circles) and from systems without Abell 
clusters (open circles)
(in units of solar luminosity $h^{-2}~L_{\sun}$)
versus the  distances of the groups (in \Mpc\ ).}
\includegraphics{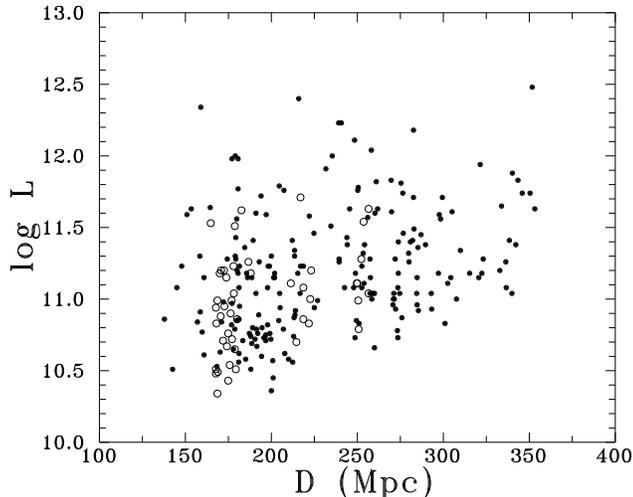}
\label{fig:scld}
\end{figure}

The LCRS observations were performed in a fixed apparent magnitude
interval and therefore galaxies fainter or
brighter than the survey limits are not included in the survey. Thus
groups consisting of faint galaxies can be detected
 only in the nearest regions
of the survey. With increasing distance, groups containing fainter
galaxies gradually disappear from the sample.  This effect 
was discussed in detail in 
Einasto et al. (\cite{env}) and in Hein\"am\"aki et al. (\cite{hei:hei}).
In Fig.~\ref{fig:scld} we show the distance-dependent selection effect
for loose groups from different systems. Fig.~\ref{fig:scld} 
shows that this distance-dependent selection effect 
affects loose groups from different systems 
in a similar way. 

\begin{figure}
\vspace{7cm}
\caption{Maximum relative densities in superclusters 
(Table~1) versus  the distance  of 
a supercluster. Filled circles correspond to superclusters 
of Abell clusters, open circles -- to systems without Abell 
clusters.}
\includegraphics{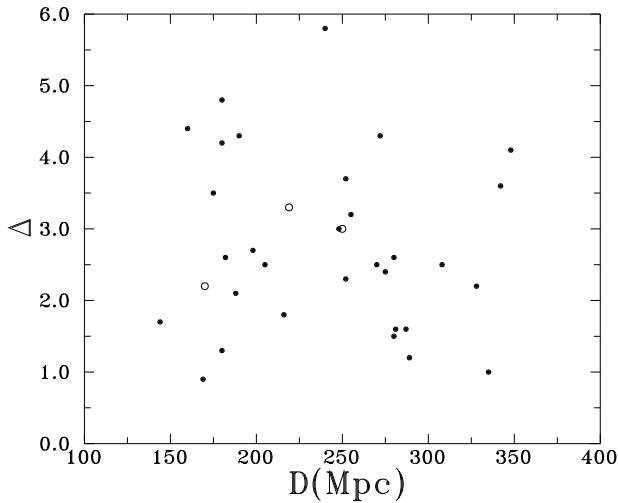}
\label{fig:dendis}
\end{figure}

In addition, in Fig.~\ref{fig:dendis} we plot the maximum relative 
supercluster densities from Table~1 against the distances 
of superclusters. This figure shows that there is no 
distance-dependent bias for supercluster densities. This 
means that the selection effects have been taken into account properly 
when determining the density field superclusters.

Einasto et al. (\cite{env}) and Hein\"am\"aki et al. 
(\cite{hei:hei}) discussed several selection effects 
that could affect the properties of loose groups. We analysed the 
properties of loose groups in high density regions around rich clusters
and showed that  selection effects cannot artificially enhance the 
properties of loose groups in high density regions. 

In Einasto et al. (\cite{env}), Einasto et al. (\cite{je2003})
and Einasto et al. (\cite{ee2003}) we discuss also several other
distance-dependent selection effects. Our analysis of  properties of
groups in a wide range of environments (not just in high density
environment of superclusters) shows that groups
of lower luminosity tend to be located in lower density environment
(as shown earlier by Lindner et al. \cite{lind95}) at all distances.

\subsection{Sizes and orientations of superclusters}
Table~1 shows that the lengths of the 
largest semiaxes of superclusters as determined by LCLGs 
are of order of $25$~\Mpc. This value is close to the minor
semiaxes of Abell superclusters determined using triaxial 
ellipsoids (Jaaniste et al.  \cite{ja98}). 
This estimate of the sizes of superclusters is in accordance with
the scales found in the power spectrum of galaxies.
At scales of about $30 - 60$~\Mpc\  the power spectrum of galaxies 
from deep surveys indicate a possible presence of a ``wiggle'' -- an 
excess near these scales and a minimum at larger scales (Silberman et 
al. \cite{sil01}, Percival et al. \cite{per01}). Gramann \& H\"utsi 
({\cite{gra01}) interpreted this scale as a scale that corresponds to  
a size of a typical supercluster (Jaaniste et al. \cite{ja98}). 

The orientation of superclusters with 
respect to the line-of-sight has an excess of 
systems oriented perpendicular to the line of sight. 
In Jaaniste et al. (\cite{jenam02}) 
we found a similar tendency of orientations in the case of
superclusters of LCLGs determined using the FoF method and 
a neighbourhood radius $R = 12$\Mpc.  
In a recent study of 
the 2dF survey Peacock et al. (\cite{pea01})
found evidence of recessional velocities
caused by a systematic infall of galaxies into superclusters. Our results 
on the orientations of superclusters may be evidence 
of the same effect.

Moreover, the supercluster SCL126 in the  direction of the Virgo 
constellation can be interpreted as an example of such infall. 
In this case
the ellipsoids calculated using data on Abell 
clusters, loose groups and individual galaxies were shown above 
(Fig.~\ref{fig:s126e}). In all cases the ellipsoid with the axes ratio about 
1:4 is located perpendicularly to the line of sight. 
Jaaniste et al. (\cite{ja98}) found that this is one of the 
flattest and thinnest superclusters, 
being located almost perpendicularly with respect to 
the line of sight. This may be an evidence of the ``squashing effect'' of 
of infalling galaxies into superclusters before 
turnaround or beginning of the relaxation (Kaiser 
\cite{kai87}), accompanied by merging and other processes that cause 
X-ray and radio radiation from clusters in this supercluster (Sect. 4).

\subsection{Individual superclusters}
The LCRS slices are, in general, too thin to contain completely 
a whole supercluster, as traced by Abell clusters. 
However, the core region of the supercluster SCL126 is located
in the slice $\delta= -3^{\circ}$. In this region the concentration 
of Abell and X-ray clusters and Las Campanas 
loose groups and galaxies in very high. 
Superclusters like SCL126 may be detected 
though the Sunyaev-Zeldovich effect in 
surveys like the forthcoming Planck mission. 

Such a high concentration of clusters has been 
observed so far only in a very few superclusters. Among them 
are the Shapley supercluster (Bardelli et al. \cite{bar00}) and
the Aquarius supercluster (Caretta et al. \cite{car02}). A very 
small number of such a high density cores of superclusters is
consistent with the results from
N-body calculations which show that such high density regions 
(the cores of superclusters that may have
started the collapse) are rare (Gramann \& Suhhonenko \cite{gra02}). 

In the case of the Horologium-Reticulum supercluster (SCL48)
the LCRS data trace rather well two of the three 
concentrations of galaxies determined in this supercluster, 
using data on rich clusters (see also Rose et al. \cite {ros02}).

The supercluster SCL100 is also located almost inside the slice 
$\delta= -3^{\circ}$. However, in contrast to SCL126, 
the local density in the region of this supercluster
is not high. With its elongated shape this supercluster 
resembles a filament. Such long filament-like superclusters 
have also been detected in $N$-body calculations (Faltenbacher 
et al. (\cite{fal:fal}).

\subsection{Large scale structure}
In Jaaniste et al. (\cite{jenam02}) we compared the smoothed 
density field of  the Las Campanas Redshift survey
calculated by Einasto et al. (in preparation) using data
on galaxies, with intersections of supercluster 
ellipsoids in the same regions. The intersection ellipses 
coincided well with the densest regions of the density field. 
In other words, the superclusters can be well traced using both 
the LCRS data and the Abell cluster data. 

We note that the LCRS slices cross the supercluster-void network at 
such an angle that the $120$ \Mpc\ scale that characterises the 
distribution of rich clusters and superclusters (Einasto et al. 
\cite{e1994}) is 
not clearly expressed, and we see a smaller scale (of about $100$ 
\Mpc) as an excess in the correlation function of LCRS 
(Tucker et al. \cite{tuc97}) instead. 

\subsection{Properties of groups in superclusters}
We  showed that loose groups in Abell superclusters
are more massive and luminous and have larger velocity
dispersions than loose groups in systems without rich clusters. 
Our
study extends the environmental enhancement of the mass and 
richness of loose groups in the vicinity of rich clusters of galaxies
described in Einasto et al. (\cite{env}) to larger scales
up to about $15 - 20$\Mpc\ from rich
clusters in superclusters. 

These results are in accordance with those by Einasto et al. 
(\cite{je2003}) who  used a larger sample of groups and clusters from the 
Sloan survey to show that groups and clusters in high density 
environments have higher luminosities than those in low density 
environments. 

Our results describe one aspect of the hierarchy of systems in the
Universe, earlier  characterised using the
sizes of voids determined by different objects (Lindner et
al. \cite{lind95}, Arbabi-Bidgoli \& M\"uller \cite{arb01}).

Several recent studies of the correlation function of nearby groups of 
galaxies show that properties of groups of galaxies in high 
density regions are different from properties of groups on average 
(Giuricin et al. \cite{giu01}, Girardi et al. \cite{gir00}, and 
Merchan et al. \cite{mer00}). Stronger clustering is an indication that 
these groups could be located in the high density regions of 
superclusters (Einasto et al. \cite{e97}, Tago et al. \cite{tag02}). 

Additionally, several studies of clusters of galaxies have provided 
evidence that properties of rich clusters depend on their large 
scale environment (Einasto et al. \cite{e2001}, Plionis \& 
Basilakos \cite{plb02}, Schuecker et al. \cite{schu01},  Chambers
et al. \cite{cham01} and Novikov et al. \cite{nov99}) up to a 
distance of about  $20$~\Mpc. This distance is close to the so-called 
``pancake scale'' (Melott \& Shandarin \cite{mel93}), 
and corresponds to the mean thickness of superclusters (Einasto et al. 
\cite{e1994} and \cite{e1997}, and Jaaniste et al. \cite{ja98}).
This distance 
is also close to that up to which the environmental enhancement of loose 
groups hase been detected in the present study.

Suhhonenko (\cite{suhh}) has demonstrated using different N-body
simulations that in simulated superclusters more massive clusters are
located in the central regions of superclusters.

Gottl\"ober et al. (\cite{got:got}) 
and Faltenbacher et al. (\cite{fal:fal})
 analysed high-resolution simulations of  
formation of galaxies, groups, and clusters and found a significant 
enhancement of the mass of haloes in the environment of other 
haloes. This effect is especially significant at scales below 10 
$h^{-1}$Mpc. Therefore, environmental enhancement of the halo mass 
is a direct evidence for the process of the hierarchical formation of 
galaxy and cluster haloes.

\section{Conclusions} 

We studied the Las Campanas loose groups in superclusters of Abell clusters. 
We described the superclusters that are crossed by LCRS slices, and the 
large-scale distribution of the  LCLGs in the supercluster-void network. 

Our results show that the orientation of superclusters, 
as determined by LCLGs, 
 has an excess of systems oriented 
perpendicularly to the line-of-sight. The Las Campanas loose groups 
in superclusters are richer and more massive than loose groups in 
systems that do not belong to superclusters. The data about galaxies, 
loose groups and rich clusters show that the 
supercluster SCL126 has a very high density core containing  
several X-ray clusters. This supercluster is located almost 
perpendicularly in respect to the line-of-sight. We assume that this may be 
due to infall of galaxies into the supercluster.

Our study indicates the importance of the role of superclusters
as high density environment which affects the properties 
(formation and evolution) of galaxy systems.

\begin{acknowledgements} 

We  thank Erik  Tago  and Heinz  Andernach  for providing us  with the
compilation of the  data on Abell  clusters. We thank Enn Saar  and
Sahar   Allam  for stimulating discussions.    The  present study  was
supported by the Estonian Science Foundation  grant 4695 and  by 
the Estonian
Research and Development  Council   grant  TO 0060058S98.   P.H.   was
supported  by   the   Finnish Academy  of   Sciences    (grant 46733).
D.L.T. was supported by  the  US Department  of Energy under  contract
No.  DE-AC02-76CH03000.  This study  has made   use  of the  NASA/IPAC
Extragalactic Database (NED) which  is operated by the Jet  Propulsion
Laboratory, Caltech, under agreement with the National Aeronautics and
Space Association.

\end{acknowledgements}


\begin{thebibliography}{}

\bibitem[1958]{abell} Abell, G., 1958, ApJS 3, 211

\bibitem[1989]{aco}  Abell, G., Corwin, H. \& Olowin, R., 1989, ApJS
70, 1 

\bibitem[1998]{at98} Andernach, H. \& Tago, E., 1998, in {\em Large
Scale Structure: Tracks and Traces}, eds.\ V.~M\"uller,
S.\,Gottl\"ober, J.P.\,M\"ucket \& J.\,Wambsganss, World Scientific,
Singapore, p.\ 147

\bibitem[2002]{arb01} Arbabi-Bidgoli, S., \& M\"uller, V.,
2002, MNRAS, 332, 205

\bibitem[2000]{bar00} Bardelli, S., Zucca, E., Zamorani, G., Moscardini, L., 
\& Scaramella, R., 2000, MNRAS, 312, 540


\bibitem[2001]{bpr01} Basilakos, S., Plionis, M., \& Rowan-Robinson,
M., 2001, MNRAS,  323, 47 

\bibitem[2002]{car02} Caretta, C.A., Maia, A.G., Kawasaki, W., 
\& Willmer, C.N.A. 2002, AJ, 123, 1200

\bibitem[2002]{cham01} Chambers, S.W., Melott, A.L., \& Miller, C.J., 
2002, ApJ, 565, 849

\bibitem[2001]{col01} Colless, M., Dalton, G., Maddox, S., 
Sutherland, W., et al. 2001, MNRAS, 328, 1039

\bibitem[1999]{deg:deg} De Grandi, S., 
B\"ohringer, H., Guzzo, L., et al. 1999, ApJ, 514, 148

\bibitem[2001]{don} Donelly, R. H., Forman, W., Jones, C., 
et al. 2001, ApJ, 562, 254

\bibitem[1997]{e97} Einasto, J., Einasto, M., Frisch, P., et al.  
1997, MNRAS, 289, 801 

\bibitem[2003b]{je2003} Einasto, J., H\"utsi, G., Einasto, M., 
et al. 2003b, AA, accepted [astro-ph/0212312]

\bibitem[2003c]{ee2003} Einasto, J., Einasto, M., H\"utsi, G.
et al. 2003c, AA, submitted 


\bibitem[1994]{e1994} Einasto M., Einasto J., Tago, E., Dalton, G. \&
Andernach, H., 1994, MNRAS, 269, 301 

\bibitem[2001]{e2001} Einasto, M., Einasto, J., Tago, E., 
Andernach, H. \& Dalton, G., 2001, AJ, 122, 2222

\bibitem[2003a]{env} Einasto, M., Einasto, J., 
M\"uller, V., Hein\"am\"aki, P.,  \&  Tucker, D.L., 2003a, 
AA, 401, 851


\bibitem[2002b]{e2002} Einasto, M., Einasto, J., Tago, E., 
et al. 2002, AJ, 123, 51


\bibitem[1997]{e1997} Einasto, M., Tago, E., Jaaniste, J., Einasto, J.
\& Andernach, H., 1997, AA Suppl. 123, 119 

\bibitem[2002]{fal:fal} Faltenbacher, A., Gottl\"ober, S., Kerscher, M., 
\& M\"uller, V., 2002, AA, 295, 1

\bibitem[1995] {fri95} Frisch, P., 
Einasto, J., Einasto, M., et al. 1995, AA 296, 611.

\bibitem[2000]{gir00} Girardi, M., Boschin, W., \& da Costa, L.N. 2000, 
A\&A, 353, 57

\bibitem[2001]{giu01} Giuricin, G., Samurovic, S., Girardi, M., 
Mezzetti, M., \& Marinoni, C. 2001, ApJ, 554, 857


\bibitem[2002]{got:got} Gottl\"ober, S., Kerscher, M., Klypin, A., 
Kravtsov, A., \& M\"uller, V., 2002, A\&A, 387, 778 

\bibitem[2002]{gra02} Gramann, M. \& Suhhonenko, I. 2002,
MNRAS, 337, 1417

\bibitem[2001]{gra01} Gramann, M., \& H\"utsi, G., 2001, 
MNRAS, 327, 538


\bibitem[2003]{hei:hei} Hein\"am\"aki, P.,  Saar, E., Einasto,  J.,  
Einasto, M., \& Tucker, D., 2003, AA, 397, 63 

\bibitem[1998]{ja98} Jaaniste, J., Tago, E., Einasto, M., et al. 
1998, AA 336, 35

\bibitem[2002]{jenam02}Jaaniste, J., Einasto, M., \& Einasto, J., 2002, 
Deep slices and the Supercluster-void Network, ed. Barbosa, D., 
JENAM 2002 (in press)


\bibitem[1978]{joe78} J\~oeveer, M., \& Einasto, J., 1978,
in {\it The Large Scale Structure of the Universe}, p. 241, eds. Longair, 
M.S., \& Einasto, J., Reidel, Dordrecht, Holland 

\bibitem[2002]{kol02} Kolokotronis, V., Basilakos, S.,  \& 
Plionis, M., 2002, MNRAS,  331, 1020 

\bibitem[1987]{kai87} Kaiser, N. 1987, MNRAS 227, 1

\bibitem[1961]{korn61} Korn, G.A., \& Korn, T.M:, 1961, {\it Mathematical
Handbook for Scientists and Engineers}, McGraw-Hill Book Co., Inc.

\bibitem[1999]{kull99} Kull, A., \& B\"ohringer, H., 1999, AA 341, 23


1996, ApJ 456, L1

\bibitem[1995]{lo95} Ledlow, M.J., \& Owen, F., 1995, AJ 109, 853



\bibitem[1995]{lind95} Lindner, U., Einasto, J., Einasto, M., et al. 
1995, AA, 301, 329



\bibitem[1993]{mel93} Melott, A.L., \& Shandarin, S.F., 
1993, ApJ, 410, 496.

\bibitem[2000]{mer00} Merchan, M., Maia, M.A.G., \& Lambas, D.G. 
2000, ApJ, 545, 26

\bibitem[1999]{nov99} Novikov, D.I., Melott, A.L., 
Wilhite, B.B., et al. 1999, MNRAS, 304, L5

\bibitem[2001]{pea01} Peacock, J.A., Cole, S., Norberg, P., 
Baugh, C.M., et al., 2001, Nature, 410, 169

\bibitem[2001]{per01} Percival, W.J., Baugh, C.M., Bland-Hawthorn, J., 
et al., 2001, MNRAS, 327, 1297


\bibitem[2002]{plb02} Plionis, M., \& Basilakos, S., 2002, 
MNRAS, 329, L47 

\bibitem[1992]{pli92} Plionis, M., Valdarnini, R., \& 
Jing, 1992, ApJ 398, 12

\bibitem[2001]{rin01} 
Rines, K., Mahdavi, A., Geller, M.J., et al. 2001, ApJ, 555, 558 


\bibitem[2002]{ros02} Rose, J.A., Gaba, A.E., Christiansen, W.A.,
et al., 2002, AJ 123, 1216

\bibitem[2001]{schu01} Schuecker, P., Boehringer, H.,  
Reiprich, T.H., Feretti, L., 2001, AA, 378, 408

\bibitem[2000]{sch:sch} Schwope, A.D., Hasinger, G., 
Lehmann, I., et al. 2000, Astron. Nachr., 321, 1 (RBS)


\bibitem [1996] {she:she} Shectman, S. , Landy, S., Oemler, A., 
et al. 1996, ApJ, 470, 122

\bibitem[2001]{sil01} Silberman, L., Dekel, A., Eldar, A., \& 
Zehavi, I. 2001, ApJ,  557, 102

\bibitem[2002]{suhh} Suhhonenko, I., 2002, PhD Thesis, 
Tartu University

\bibitem[2002] {tag02} Tago, E., Saar, E., Einasto, J., et al.
 2002, AJ, 123, 37 

\bibitem[1978]{tar78} Tarenghi, M., Tifft, W., Chincarini, G., 
Rood, H., \& Thompson, L., 1978,
in {\it The Large Scale Structure of the Universe}, p. 241, eds. Longair, 
M.S., \& Einasto, J., Reidel, Dordrecht, Holland 

\bibitem [2000] {tuc:tuc} Tucker, D.L., Oemler, A.Jr., Hashimoto, Y.,
et al. 2000, ApJS, 130, 237

\bibitem[1997]{tuc97} Tucker, D. L., Oemler, A., Jr.,
 Kirshner, R. P., et al. 1997, MNRAS 285, L5

\bibitem[1997]{vet97} Vettolani, G., Zucca, E.,
Zamorani, G., et al. 1997, AA 325, 954

\bibitem[1989]{w89} West, M. 1989, ApJ 347, 610

\bibitem[2000]{yo0} York, D.G., Adelman, J., Anderson, J.E., 
et al. 2000, AJ, 120, 1579 


\bibitem[1982]{zes82}  Zeldovich, Ya.B., Einasto, J. \&
Shandarin, S.F.   1982, Nature 300, 407


\end{thebibliography}
\end{document}